\documentclass[a4 paper, 12 pt] {article}

\usepackage{graphicx}
\usepackage{amsmath}
\usepackage{amssymb}
\usepackage{bm}

\def\be{\begin{equation}}
\def\ee{\end{equation}}
\def\bdi{\begin{displaymath}}
\def\edi{\end{displaymath}}
\def\br{\begin{eqnarray}}
\def\er{\end{eqnarray}}

\def\RR{{\rm I\kern-.1567em R}}                              % Doppel R 
 \def\CC{{\rm C\kern-4.7pt                                    % Doppel C 
 \vrule height 7.7pt width 0.4pt depth -0.5pt \phantom {.}}} 
 \def\ZZ{{\sf Z\kern-4.5pt Z}}                                % Doppel Z 

\begin{document}

\begin{titlepage}
\vspace*{-2 cm}
\noindent

\vskip 3cm
\begin{center}
{\Large\bf Comment on ``Reduction of static field equation of Faddeev 
model to first order PDE'', 
%%PLB 652 (2007) 384, 
arXiv:0707.2207  }
\vglue 1  true cm

C. Adam$^{a*}$,   J. S\'anchez-Guill\'en$^{a**}$,  
and A. Wereszczy\'nski$^{b\dagger}$
\vspace{1 cm}

\small{ $^{a)}$Departamento de Fisica de Particulas, Universidad
      de Santiago}
      \\
      \small{ and Instituto Galego de Fisica de Altas Enerxias (IGFAE)}
     \\ \small{E-15782 Santiago de Compostela, Spain}
      \\ \small{ $^{b)}$Institute of Physics,  Jagiellonian
     University,}
     \\ \small{ Reymonta 4, 30-059 Krak\'{o}w, Poland}

\medskip
\end{center}

\normalsize
\vskip 0.2cm

\begin{abstract}
The authors of the article Phys. Lett. B 652 (2007) 384, 
\newline (arXiv:0707.2207)
propose an interesting 
method to solve the Faddeev model by reducing it to a set of first
order PDEs. They first construct a vectorial quantity $\bm \alpha $, depending
on the original field and its first derivatives, in terms of which the field
equations reduce to a linear first order equation. Then they find vectors
$\bm \alpha_1$ and  $\bm \alpha_2$ which identically obey this linear
first order equation. The last step consists in the identification of
the $\bm \alpha_i$ with the original $\bm \alpha$ as a function of the
original field. 
Unfortunately, the derivation of this last step in the paper cited above
contains an error which invalidates most of its results.

\end{abstract}

\vfill

{\footnotesize
$^*$adam@fpaxp1.usc.es

$^{**}$joaquin@fpaxp1.usc.es

$^{\dagger}$wereszczynski@th.if.uj.edu.pl }

\end{titlepage}

%\section{Introduction}

The Faddeev model \cite{Fad}, \cite{FN1}
(also known as the Skyrme--Faddeev model or the 
Faddeev--Niemi model) is a nonlinear field theory in 3+1 dimensions
which is known to support knotted solitons, both from an analysis of its
topology and stability \cite{LiYa}, and from numerical calculations
\cite{GH} - \cite{HiSa}. Apart from their
existence, however, the analytic information on these solitons is rather
sparse. 

In the letter \cite{HS}, the authors proposed a method to partially solve the
static field equations by effectively reducing them to a set of first order
equations. Unfortunately, that paper contains an error which
invalidates most of its results. In the sequel we briefly review the
construction of \cite{HS}, point out the error and demonstrate that from their
(incorrect)
results, incorrect conclusions may be drawn (i.e., one may 
construct ``solutions'' which are well-known {\em not} to be solutions of the
Faddeev model). 

The target space of the Faddeev model is the two-sphere and may be
described either by a three-component unit vector field $\vec n$ or
by a complex field $u$ via stereographic projection. 
The energy functional for static configurations of the Faddeev model 
(in terms of the complex field $u$) is
\be
E[u,\bar u ] = \int d^3 {\bf x} (c_2 \epsilon_2 + c_4 \epsilon_4 )
\ee
with
\begin{equation}
\epsilon_2=\frac{4}{(1+|u|^2)^2}(\nabla u\cdot \nabla u^{*}),
\end{equation}
\be
\epsilon_4=-8\frac{( \nabla u\times \nabla u^*)^2}{(1+|u|^2)^4}.
\ee
Following the conventions of \cite{HS}, we now assume a choice of length 
units such that $c_2 =4c_4$ and re-express $u$ by its modulus and phase,
\be
u=R e^{i\Phi}
\ee
with real functions $R$ and $\Phi$. Then the static field equations can be
written like
\be \label{HS1}
\nabla \cdot {\bm \alpha} + i{\bm \beta} \cdot {\bm \alpha} =0
\ee
and its complex conjugate, where
\be
{\bm \alpha} \equiv \frac{\nabla u^*}{1+R^2 } - \frac{\nabla u^* 
\times (\nabla u \times \nabla u^* )}{(1+R^2)^3}
\ee
and
\be
{\bm \beta} \equiv -i \frac{u^* \nabla u - u \nabla u^* }{1+R^2}
= \frac{2R^2}{1+R^2} \nabla \Phi .
\ee
Equation (\ref{HS1}) is the starting point for the analysis in Ref.
\cite{HS}. Next, the authors observe that the vectors
\br 
{\bm \alpha}_1 &=& (\nabla R \times \nabla \rho )
\exp \left( -2i \Phi \frac{R^2 }{1+R^2} \right) \nonumber \\
{\bm \alpha}_2 &=& (\nabla \Phi \times \nabla \mu ) \label{alpha_i}
\er
identically obey Eq. (\ref{HS1}) for arbitrary complex functions $\rho$ and
$\mu$. Due to the linearity of Eq. (\ref{HS1}), also the sum ${\bm \alpha }_1
+  {\bm \alpha }_2$ obeys this equation.  

For a further analysis, the authors then regard $\rho$ and $\mu$ as
functions of $R$, $\Phi$ and a third function $\zeta$ which is unknown at this
moment but should obey
$\frac{\partial(R, \Phi,\zeta)}{\partial(x_1,x_2,x_3)}\neq 0$ such that 
the three functions $R$, $\Phi$, $\zeta$ may be used as a new system of
curvilinear coordinates. The idea is then to expand the vectors 
${\bm \alpha }$ and ${\bm \alpha }_i$ into the basis
\be \label{basis}
\nabla R \, ,\quad R\nabla \Phi \, , \quad R\nabla R\times \nabla \Phi
\ee
and to compare coefficients. For the gradient of $\zeta$ the authors assume
\begin{equation} \label{grad}
\nabla \zeta=\gamma \nabla R\times R\nabla \Phi+\xi \nabla R+R\eta\nabla \Phi.
\end{equation}
where $\gamma, \xi$ and $\eta$ are, at this moment, {\em unconstrained}
real functions. This assumption is the  error we announced at the
beginning. The l.h.s. of Eq. (\ref{grad}) is a gradient and, therefore, obeys
$\nabla \times \nabla \zeta =0$. Applying this condition to the r.h.s.
of the same equation produces constraints which the functions 
$\gamma, \xi$ and $\eta$ have to obey. Concretely, in an index notation
the constraints are
\be
\left( \gamma R R_k \Phi_j \right)_j - \left( \gamma RR_j \Phi_k \right)_j
+ \epsilon_{kjl}\left( \xi_j R_l + R \eta_j \Phi_l + \eta R_j \Phi_l \right)
=0
\ee
where the subindices mean partial derivatives. Obviously, the constraints
contain first derivatives of the functions $\gamma, \xi$, $\eta$, as well as
second derivatives of $R$ and $\Phi$, and it is not known how to expand these
expressions
into the basis (\ref{basis}). This problem invalidates all the subsequent
analysis of Ref. \cite{HS}, where the comparison of ${\bm \alpha }$ with
${\bm \alpha}_1 + {\bm \alpha}_2$ essentially leads to a system of
linear equations. 

Let us illustrate how the results of Ref. \cite{HS} lead to wrong conclusions,
by showing that using these results one may derive easily 
%%Finally, let us demonstrate that by using nevertheless the results of  
%%Ref. \cite{HS}, it is easy to derive 
``solutions'' of the Faddeev model
which are well-known not to be solutions at all.
For this purpose, we first summarize the (incorrect) final result of Ref.
\cite{HS}. The result essentially says that there are six real functions
(the three functions $\gamma, \xi$ and $\eta$, as well as three more functions
called
$a$, $b$ and $c$, which are related to the arbitrary complex functions 
$\rho$ and $\mu$ of Eq. (\ref{alpha_i})), which have to obey a system of
two linear first order PDEs (Eq. (54) of Ref. \cite{HS}). Any choice of
these six functions obeying the two linear first order PDEs automatically
provides a static solution for the Faddeev model. More precisely, it directly
provides a solution for the three quantities
\br
p&=&(\nabla R)^2=S(R,\Phi,\zeta)  ,\\
q&=&\nabla R\cdot R\nabla \Phi=T(R,\Phi,\zeta) ,\\
r&=&(R\nabla \Phi)^2= U(R,\Phi,\zeta),
\er  
(i.e., it provides the r.h.s. of these equations),
from which $R$ and $\Phi$ still have to be calculated. 

Now, in order to find some specific solutions, 
let us make some simplifying assumptions for the functions
$a$, $b$, $c$. Concretely, we assume $a=0$ and $b=c$,  
which immediately leads to
\br
p=r &=& [(b(1+R^2)]^{-1}  ,\\
q&=& 0 ,
\er
see Eqs. (39)-(41) of Ref. \cite{HS}. Further, the system of
two linear first order PDEs (Eq. (54) of Ref. \cite{HS}) decouples under these
assumptions. Next, we make the further assumption that $\gamma =$ const.,
then the l.h.s. of Eq. (54) of Ref. \cite{HS} is zero. The resulting
two first order differential equations are now ordinary ones and are just the
defining equations for the (up to now, arbitrary) functions
$\xi$ and $\eta$, respectively,
for a given but completely arbitrary
function $b$. This implies that {\em any} solution to the equations
\br 
 (\nabla R)^2 &=&  (R\nabla \Phi)^2 \\
 \nabla R\cdot R\nabla \Phi &=&0
\er
(the so-called complex eikonal equation)
should be a solution to the field equations of the Faddeev model (due to the
arbitrariness of the function $b$). But this conclusion is certainly wrong.
It is, for instance, well-known that the ansatz in toroidal coordinates
\be \label{tor-ans}
u=f(\tilde \eta ) e^{i n\tilde \xi + i m \tilde \varphi }
\ee
provides solutions to the complex eikonal equation for arbitrary integers
$m$ and $n$, see \cite{eik}, \cite{eik2} 
(we use tildes for the torus coordinates in order
not to confuse them with the functions introduced above; for the conventions
used for the
 torus coordinates, we refer, e.g., to \cite{eik}). 

On the other hand, it is well-known that the ansatz (\ref{tor-ans}) in
toroidal coordinates is {\em incompatible} with the field equations of the
Faddeev model, see, e.g., \cite{Kundu}. 

In short, we have demonstrated that the analysis of Ref. \cite{HS}
contains an error, and that the use of the (incorrect) results of that paper
may lead to wrong conclusions about solutions of the Faddeev
model, which was the purpose of this comment. 

We think, nevertheless, that the starting point of the paper  \cite{HS},
i.e., the linear equation (\ref{HS1}) and the observation that it is
identically obeyed by the family of vectors of Eq. (\ref{alpha_i}), is
interesting and deserves further investigation.

\section*{Acknowledgement}
 A.W. gratefully acknowledges support from Adam
Krzy\.{z}anowski Fund and Jagiellonian University (grant WRBW
41/07). C.A. and J.S.-G. thank MCyT (Spain) and FEDER
(FPA2005-01963), and 
Xunta de Galicia (grant PGIDIT 06 PXIB 296182 PR and Conselleria de
Educacion) for support. 
Further, C.A. acknowledges support from the Austrian
START award project FWF-Y-137-TEC and from the FWF project P161 05
NO 5 of N.J. Mauser.

\end{document}